\documentclass[twocolumn]{aastex702}

\usepackage{amsmath}
\usepackage{booktabs}
\usepackage{xcolor}

\newcommand{\ddc}{\textsc{DDC}}
\newcommand{\kpolaris}{\textsc{KPolaris}}
\newcommand{\ipole}{\texttt{ipole}}

\shorttitle{Dense Dump Codec}
\shortauthors{Zhang et al.}

\begin{document}

\title{Dense Dump Codec: Error-Controlled, Random-Access Compression of
GRMHD Time Series for Slow-Light Radiative Transfer}

\author[orcid=0009-0008-9585-1807]{Zelin Zhang}
\affiliation{Institute of Fundamental Physics and Quantum Technology, \& School of Physical Science and Technology, Ningbo University, Ningbo, Zhejiang 315211, P. R. China}
\affiliation{Zhejiang Key Laboratory of Extreme Universe, \& BINGO Center, Ningbo University, Ningbo, Zhejiang 315211, China }
\email{zhangzelin1@nbu.edu.cn}

\author[orcid=0000-0003-0869-4601]{Zhenyu Zhang}
\affiliation{Institute of Fundamental Physics and Quantum Technology, \& School of Physical Science and Technology, Ningbo University, Ningbo, Zhejiang 315211, P. R. China}
\affiliation{Zhejiang Key Laboratory of Extreme Universe, \& BINGO Center, Ningbo University, Ningbo, Zhejiang 315211, China }
\email{zhangzhenyu@nbu.edu.cn}

\author[orcid=0000-0003-4509-9705]{Bin Chen}
\email{chenbin1@nbu.edu.cn}
\affiliation{Institute of Fundamental Physics and Quantum Technology, \& School of Physical Science and Technology, Ningbo University, Ningbo, Zhejiang 315211, P. R. China}
\affiliation{Zhejiang Key Laboratory of Extreme Universe, \& BINGO Center, Ningbo University, Ningbo, Zhejiang 315211, China }
\affiliation{School of Physics, \& Center for High Energy Physics, Peking University, No.5 Yiheyuan Rd, Beijing 100871, P. R. China}
\correspondingauthor{Bin Chen}

\begin{abstract}
Black-hole movies provide a unique venue to study the time evolution of accreting plasma. Modeling this evolution with slow-light radiative transfer requires closely spaced simulation outputs, because the plasma changes as light propagates through it. Saving these outputs in full is costly, whereas sparse sampling introduces time-interpolation errors.
We present the Dense Dump Codec (DDC), a compression method for dense GRMHD time series. DDC stores exact anchor states and compactly represents the intermediate evolution, while allowing direct access to selected times and variables. 
Applied to eight production sequences spanning SANE and MAD simulations at several black-hole spins with $0.1M$ output, DDC archives are about 10 times smaller than the original dense data and even about 50\% smaller than the original $0.5M$ output in aggregate.
We further integrate a native DDC reader into polarized slow-light radiative transfer. Tests at 86, 230, and 345~GHz show that DDC may reduce aggregate Stokes-image errors by 54--79\% relative to slow-light calculations using original $0.5M$ input. These results indicate that DDC enables high-cadence slow-light calculations without the prohibitive cost of saving every dense GRMHD state in its original form.

\end{abstract}


\section{Introduction}
\label{sec:introduction}

Horizon-scale observations provide a direct view of the structure and dynamics of black-hole accretion flows. The Event Horizon Telescope (EHT) has resolved the emission surrounding M87* and Sgr~A* \citep{eht2019m87l1,eht2022sgra1}, and its observations of Sgr~A* reveal structural variability beyond that expected from measurement uncertainties or interstellar scattering \citep{eht2022sgra4}.
This variability supplies information beyond the time-averaged image, with light-curve measurements already placing strong constraints on accretion-flow models \citep{eht2022sgra5}. Planned extensions through the next-generation EHT aim to follow source evolution with improved imaging capabilities \citep{johnson2023ngeht}. These developments motivate synthetic black-hole movies that connect the observed changes in emission to the evolving accretion flow.

Modeling these movies requires accounting for light-travel-time effects when the source varies on timescales comparable to the propagation delays across the image. 
In rapidly evolving sources, the plasma can change appreciably during the time it takes light to cross the emitting region.
Relativistically moving features in the jet-launching region provide an example in which this evolution alters the observed morphology and variability \citep{tsunetoe2026slowlight}. Strong lensing provides another route: photons that follow longer paths around the black hole can arrive substantially later than those reaching the observer more directly, so different image components sample different stages of the source evolution. The resulting correlations between brightness fluctuations at different positions and observer times are particularly relevant to autocorrelation studies of black-hole movies \citep{hadar2021autocorrelations,hadar2023spectrotemporal,Zhang:2025vyx,Zhu:2025jqh,bezdekova2026correlations}. Interpreting these temporal signatures therefore requires treating the source evolution and light propagation together.

Slow-light general-relativistic radiative transfer (GRRT) incorporates
both plasma evolution and light-travel-time delays by tracing radiation
through a time sequence of GRMHD simulation states.
In the fast-light approximation, each ray propagates through a single frozen simulation state. Slow-light transfer instead evaluates the fluid at the coordinate time reached along the ray, so one image combines emission from a range of retarded times \citep{bronzwaer2018raptor,moscibrodzka2018ipole,white2022blacklight,Zhu:2026byh}.
The resulting changes depend on the source, ranging from modest light-curve differences to substantial changes in morphology and variability \citep{bronzwaer2018raptor,tsunetoe2026slowlight}.
Resolving this evolution requires sufficiently close input snapshots. Interpolation between widely separated simulation snapshots can miss variations in the plasma and introduce errors in the radiative coefficients and resulting images. More closely spaced snapshots allow the transfer calculation to follow this evolution more faithfully.

Retaining the dense simulation output needed for slow light can
impose a substantial storage cost. Each saved state contains
three-dimensional fluid and magnetic-field arrays, and a single image
may sample these fields over tens to hundreds of gravitational times.
Long image sequences extend the required interval further, multiplying
the volume of data that must remain available. Saving fewer states
reduces this cost but leaves more of the intervening evolution to be
reconstructed by interpolation. Compression offers another possibility.
Retain the closely spaced states calculated by the simulation while
reducing the storage required for each state.

Appropriate compression methods can reduce the storage required for high-cadence GRMHD output. For slow-light radiative transfer, compression must preserve the plasma properties needed to calculate the radiation and allow efficient access to snapshots at the times sampled along each ray.
General compressors such as ZFP and SZ provide compact representations of floating-point arrays with configurable precision \citep{lindstrom2014zfp,di2016sz}. Studies of evolving simulation data also show that frequent output at reduced precision can preserve features better than discarding states under the same storage budget \citep{gong2023spatiotemporal}. For GRMHD, this motivates allocating precision among fluid variables with different dynamic ranges and radiative roles, then assessing the reconstructed emission as well as the fields themselves. The compressed representation must also allow a transfer calculation to retrieve the neighboring times and variables it needs without decoding an entire sequence. A representation that combines controlled compression errors with selective access can therefore reduce storage demands while remaining compatible with slow-light workflows.

We introduce the Dense Dump Codec (\ddc) \footnote{\url{https://github.com/zelinzh/dense-dump-codec}}, a compression method for high-cadence GRMHD time series. The method exploits temporal correlations by predicting intermediate simulation states from exact anchors and storing compressed, channel-dependent residuals.
These residuals retain information from states actually calculated by the simulation, while an indexed representation provides direct access to selected times and variables. The anchor states remain available in their original form for conventional analysis.
We evaluate the storage--accuracy trade-off using reconstructed fluid quantities and polarized slow-light images, and integrate a native reader into \kpolaris \footnote{\url{https://github.com/zelinzh/KPolaris}} to assess direct use of the compressed data in radiative transfer.

The remainder of this paper is structured as follows. In Section~\ref{sec:method}, we describe the main features and implementation of the Dense Dump Codec. In Section~\ref{sec:evaluation}, we introduce the GRMHD data sets, comparison baselines, and validation procedure. In Section~\ref{sec:results}, we present the compression performance and its accuracy in polarized slow-light calculations. Finally, we summarize our main conclusions in Section~\ref{sec:summary}. We use geometrized units $G=c=1$, with lengths and times expressed in units of the black-hole mass $M$, so that one time unit is $GM/c^3$ in physical units.

\section{Dense Dump Codec}
\label{sec:method}

\begin{figure*}[t]
\centering
  \includegraphics[width=\textwidth]{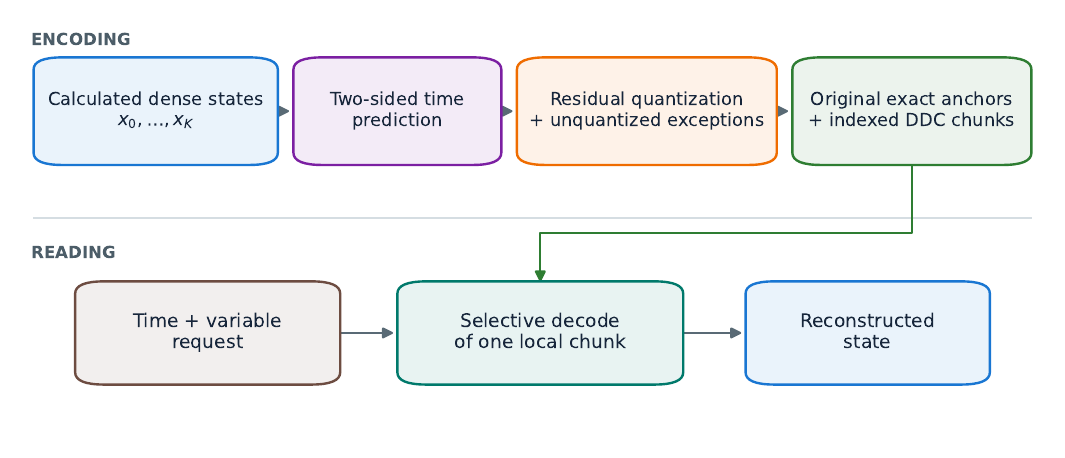}
\caption{General \ddc\ workflow. Dense states are predicted between
exact anchors, and channel residuals are stored in independently indexed
chunks. A query decodes only the anchors and local chunk needed for the
selected state and variables.}
\label{fig:architecture}
\end{figure*}

The Dense Dump Codec stores a densely sampled GRMHD time series at a small fraction of its original size while keeping every saved state directly accessible. The underlying principle is simple. Successive states of a simulation evolve smoothly, so a state can be predicted from its two exactly stored boundary states, and only the small prediction residual needs to be encoded. \ddc\ implements this principle in three stages. Intermediate states are predicted between exact anchors, the residuals are quantized per channel and per spatial block under a controlled error bound, and the results are packed into indexed containers, from which a request for any stored time and variable is answered without decoding the rest of the sequence. Figure~\ref{fig:architecture} shows the full workflow, and the subsections below describe the three stages in turn, followed by the streaming implementation.

\subsection{Physical-time prediction}

The output of a GRMHD simulation is a sequence of fluid states at successive physical times. We write $x_j$ for one array of that sequence, the values of one field or vector component on the grid at time $t_j$, and we organize the sequence into coding intervals that we call groups of pictures (GOPs), following the analogous structure in video compression, where each group is likewise bounded by independently decodable frames. Each GOP contains two exactly stored boundary states, the anchors $x_0$ and $x_K$, and $K-1$ intermediate states. An intermediate time $t$ is located inside its interval by the normalized parameter
\begin{equation}
  \tau = \frac{t-t_0}{t_K-t_0}.
  \label{eq:tau}
\end{equation}
The physical times are read from stored metadata rather than inferred from array indices, so the construction remains correct even when output times carry small numerical jitter.

For an intermediate state, the two anchors provide the prediction. Density $\rho$ and internal energy $u$ are positive over valid cells and span many orders of magnitude, so their predictor operates on logarithms and is geometric,
\begin{equation}
  \widehat{x}(\tau) =
  \exp\left[(1-\tau)\log x_0 + \tau\log x_K\right].
  \label{eq:logpredict}
\end{equation}
Velocity, magnetic field, face-centered magnetic flux, and divergence diagnostics are signed quantities, for which logarithms are not defined, so they use the linear predictor
\begin{equation}
  \widehat{x}(\tau) = (1-\tau)x_0 + \tau x_K.
  \label{eq:linpredict}
\end{equation}
With $T=\log$ for the positive thermodynamic channels and the identity for the others, the coded signal is the transformed residual
\begin{equation}
  r = T(x)-T\!\left(\widehat{x}\right).
  \label{eq:residual}
\end{equation}
The predictor absorbs the smooth evolution that the two anchors already describe, so the residual $r$ is small, and it is the only quantity that \ddc\ needs to encode for an intermediate state. The predictor does not create missing states. Every intermediate state that the codec encodes was first calculated by the simulation.

\subsection{Channel-aware residual quantization}

The codec quantizes the transformed residual $r$ defined in Equation~(\ref{eq:residual}) rather than the full fluid variable, and a decoded residual reconstructs the state as
\begin{equation}
  x_{\mathrm{dec}} =
  T^{-1}\!\left[T(\widehat{x})+\widetilde{r}\right].
  \label{eq:decode_residual}
\end{equation}
The bit depths discussed below therefore apply only to the small prediction residuals, and never to the full fluid variables or to the anchor states. The stage has two parts, a quantizer for the bulk of each block and a full-precision exception path for rare large values.

A single quantization scale would not suit every variable or every part of a GRMHD domain. Density and internal energy span many orders of magnitude, velocity and magnetic field have different typical amplitudes, and localized structures such as the funnel or current sheets can contain much larger residuals than the disk. We therefore treat each scalar field and each vector component as a separate channel, and the scale is also computed independently for each spatial block. A large residual in one region or in one magnetic-field component then does not reduce the precision used elsewhere.

Within one residual block with values $r_i$, signed $b$-bit integer codes are defined by
\begin{equation}
  q_{\max}=2^{b-1}-1 .
\end{equation}
and the quantization step is set from a robust amplitude of the block,
\begin{align}
  a_p &= \operatorname{percentile}_p\!\left(\{|r_i|\}\right), \\
  s   &= \max\!\left(\frac{a_p}{q_{\max}},s_{\min}\right), \\
  q_i &= \operatorname{round}\!\left(\frac{r_i}{s}\right).
  \label{eq:quantization}
\end{align}
where $a_p$ is the $p$th percentile of the absolute residuals and the floor $s_{\min}=10^{-30}$ prevents division by zero.

Residuals inside the integer range, $|r_i|/s\le q_{\max}$, are reconstructed as
\begin{equation}
  \widetilde{r}_i=sq_i,
  \qquad |q_i|\le q_{\max}.
\end{equation}
Rounding then bounds the transform-space error by
\begin{equation}
  |\widetilde{r}_i-r_i|\le \frac{s}{2}.
\end{equation}
This bound is subject to floating-point roundoff in the implementation. For the logarithmic channels it is a relative bound, since a transform-space error $\epsilon$ multiplies the physical value by $\exp(\epsilon)$. For the linear channels it is an absolute error on the residual.

The percentile selects a scale for the majority of the block and is not used as a clipping threshold. Residuals outside the integer range, $|r_i|/s>q_{\max}$, are therefore not clipped. The codec writes their flattened cell index and an unquantized single-precision value to a sparse exception stream, and decoding substitutes this value for the quantized code. A lower percentile sharpens the resolution of most cells while sending more values to the exception stream, and a higher percentile does the reverse. Rare large residuals can thus neither set the precision of an entire block nor be silently saturated.

Table~\ref{tab:profile} lists the channel rules used in the final experiments. Here the profile is fixed, which means that the transform $T$, the integer depth $b$, and the percentile $p$ were chosen during development and are not retuned for individual models, sequences, or windows. The numerical scale $s$ is still computed separately for every residual block. The four rows are the primitive streams used in the production and GRRT tests. Exact anchors remain unchanged, and exception values are stored without quantization regardless of the listed bit depth.

\begin{center}
\begin{minipage}{\columnwidth}
\makeatletter\def\@captype{table}\makeatother
\centering
\setlength{\tabcolsep}{5pt}
\caption{Fixed channel-dependent residual-quantization profile. Bit depths
apply only to temporal residuals.}
\label{tab:profile}
\begin{tabular}{lccc}
\toprule
Channel & Residual space $T$ & Bits $b$ & Percentile $p$ \\
\midrule
\multicolumn{4}{l}{\textit{Primitive streams}} \\
$\rho$ & logarithmic & 8 & 99.9 \\
$u$ & logarithmic & 8 & 99.9 \\
$\widetilde{u}^i$ & linear & 5 & 99.5 \\
$B^i$ & linear & 16 & 99.5 \\
\bottomrule
\end{tabular}
\end{minipage}
\end{center}

The resulting profile assigns 16 bits to $B^i$, while lower bit depths are used for the thermodynamic and velocity channels. Section~\ref{sec:results} presents the matched comparison of precision profiles.

\subsection{Indexed containers and exact anchors}

\begin{figure*}[tbp!]
\centering
\IfFileExists{figures/ddc_gop_structure.pdf}{
  \includegraphics[width=\textwidth]{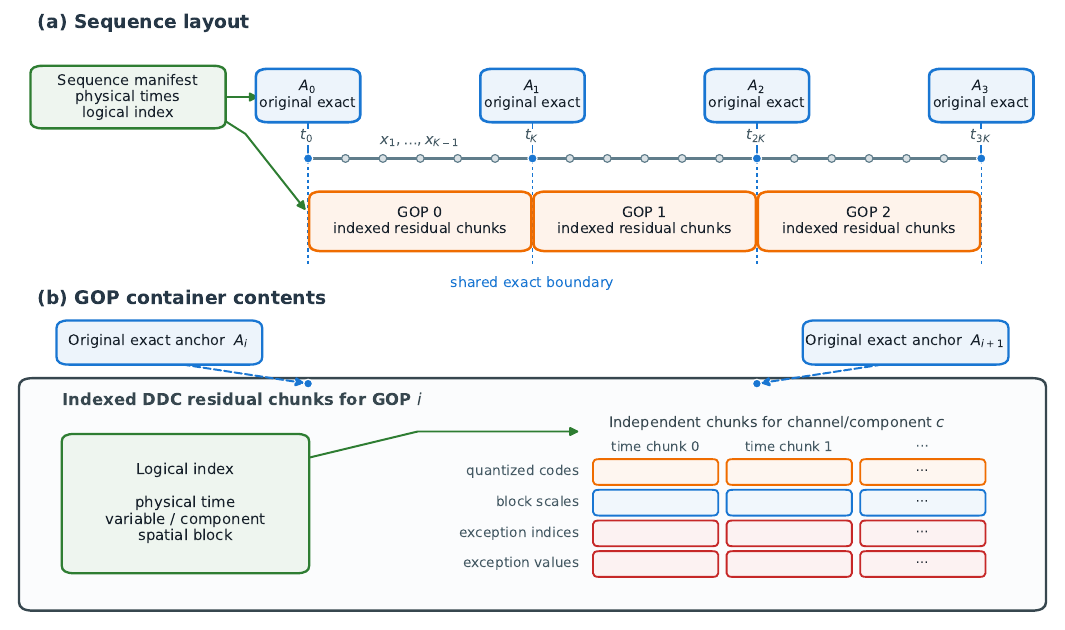}
}{
  \fbox{\parbox[c][76mm][c]{0.94\textwidth}{\centering
  Missing generated DDC package-structure figure.}}
}
\caption{Logical organization of a \ddc\ sequence. (a) Exact anchors
bound adjacent GOPs, while the manifest indexes the intermediate states shown
in gray. (b) Each GOP contains a logical index and independently decodable
channel chunks with quantized codes, scales, and unquantized exceptions.}
\label{fig:container-structure}
\end{figure*}

Figure~\ref{fig:container-structure}(a) shows how GOPs and exact anchors form a sequence. Let $A_i$ denote the exact state at one GOP boundary. In the predictor notation above, $x_0=A_i$ and $x_K=A_{i+1}$ for GOP $i$. The GOP spans this interval and represents its calculated intermediate states in an indexed \texttt{.ddc} container. The state $A_{i+1}$ is therefore both the right boundary of GOP $i$ and the left boundary of GOP $i+1$, a shared exact state rather than two separate copies. A sequence manifest maps physical times to the corresponding GOP and anchor locations.

Figure~\ref{fig:container-structure}(b) shows the contents of one GOP container. Its metadata and index members map a physical time, a variable, and a spatial block to the corresponding encoded residuals. The residuals are organized by channel and divided into short, independently decodable temporal chunks. Within a chunk, quantized codes, block scales, and sparse exceptions are packed with reversible transformations and a lossless backend, which changes only the storage size. Decoding reproduces exactly the output of the residual-quantization stage, so the only numerical error in a decoded state is the quantization error described above.

A request for one variable at one time therefore requires only the relevant local chunk and the associated anchors, rather than the complete GOP or all preceding outputs. This organization bounds both how much data a request must read and its dependence on other decoded states, while preserving direct access to stored states with nonuniform physical-time spacing.

Exact anchors are retained in their original representation. They provide the endpoints of the temporal predictor and remain directly available for conventional analysis. Because no lossy transform is applied to these boundary states, codec errors cannot propagate from one GOP into the next. A requested intermediate state therefore depends only on its two original anchors and one local residual chunk.

\subsection{Streaming creation, recovery, and random access}

A \ddc\ sequence is described by a manifest that records the physical time of each state, the available variables, the GOP containers and original exact anchors, and the information required to validate them.

In the present implementation, the archive is created asynchronously. The simulation writes dense states in its conventional output representation, and a watcher waits until every state in one GOP and its closing anchor are complete. It then encodes that interval, verifies the encoded members and the reconstructed arrays, and publishes the updated manifest atomically. Only after this commit may the transient non-anchor dense files be removed. Interrupted work resumes from the last committed GOP, and a file that is still being written is never accepted as codec input.

The codec changes only how already-calculated analysis states are retained, not the GRMHD evolution itself. The original analysis anchors remain unchanged, and \ddc\ stores the densely sampled intermediate states together with the indices needed to access them. Restart checkpoints remain independent of the codec, because they must carry the additional information required to resume the evolution and they follow the cadence chosen for simulation recovery.

For a request at a stored physical time, the decoder uses the manifest to identify the corresponding GOP, reads the two required anchors, and accesses only the chunks containing the requested variables. The decoded state can be written out in a conventional analysis-file format or returned directly to a downstream application as native arrays. Slow-light calculations typically request many nearby fluid times, so the native interface can decode GOP-aligned batches and reuse cached anchors and neighboring states. A densely sampled sequence can then be traversed without reconstructing the complete archive or loading variables that the calculation does not use.

For repeated GRRT work, a disposable spatially indexed cache lets the reader select the region requested by the transfer code. The native service prefetches neighboring chunks, sends compact residual data, and reconstructs single-precision fluid arrays on the GPU, and multiple observer times can share a time-ordered batch and its resident states. This working cache is an access optimization, and it is not part of the compact archival byte count.

\section{Evaluation}
\label{sec:evaluation}

This section sets up the evaluation. We first describe the GRMHD sequences, the codec configuration under test, and the comparison baselines. We then define the field-level accuracy and storage criteria that the compressed archive must satisfy, and finally describe the polarized slow-light calculations used to test the retained fluid data and to measure the cost of native \ddc\ access.

\subsection{GRMHD sequences and baselines}

We implemented \ddc\ in a workflow based on the open-source GRMHD code \textsc{KHARMA} \citep{prather2024kharma}, which is built on Parthenon \citep{grete2023parthenon}. The simulation runs were restarted from the public v3 GRMHD library \citep{dhruv2025survey}. 
The original representation is the parallel-HDF5 analysis output written by \textsc{KHARMA} in single precision. The primitive \texttt{prims.uvec} contains $\widetilde{u}^i$, the spatial four-velocity projected with respect to the normal observer, and \texttt{prims.B} contains the constrained-transport magnetic field $B^i$. The corresponding conserved field is $\sqrt{-g}\,B^i$.

Each state contains two native Parthenon MeshBlocks of $288\times128\times64=2{,}359{,}296$ cell-centered elements per scalar field. Vector components are separate channels within each MeshBlock, and the face-centered magnetic arrays have the corresponding
$289\times129\times65$ dimensions. The density, velocity, and magnetic channels each use a single scale per native MeshBlock. For $u$ only, each MeshBlock is subdivided into $32\times16\times8$ cells in $(r,\theta,\phi)$ before the percentile scale is computed. These 4096-cell tiles prevent large, spatially localized thermal residuals from setting the precision of the entire MeshBlock, and the $u$ bit depth remains eight. The tile layout is recorded with the residuals and does not alter the GOP predictor or the exact anchors. The original baseline uses the files as written, including their HDF5 metadata, chunking, and lossless gzip compression. We refer to this native representation as ``original
data''. It is not encoded by \ddc.

The simulation produces primitive states at a dense cadence of approximately $0.1M$, and every $2.5M$ one of these calculated states is designated as an exact analysis anchor. Two successive anchors define a GOP. The anchors are
preserved exactly, while the 24 calculated $0.1M$ states between them are
represented by their encoded residuals. Thus, the intermediate states are
not generated by interpolation between the $2.5M$ anchors. They are genuine
simulation states, stored more compactly by \ddc. The $0.1M$ state cadence and $2.5M$ anchor spacing are choices of this benchmark rather than requirements of the codec, and phrases such as ``the $0.1M$ \ddc\ sequence'' below refer to this configuration.

The evaluation uses a test set of 18 sequences restarted near $t=25000M$, $27000M$, and $29000M$. Each epoch includes SANE and MAD flows with $a_*=-0.94$, $+0.5$, and $+0.94$. The four $|a_*|=0.94$ sequences last $100M$ each, and the two $a_*=+0.5$ sequences last $300M$ each. Across the three epochs, the test set spans $3000M$ and contains $30,018$ original $0.1M$ states, including sequence endpoints. It was produced first, to test the codec and select the quantization profile, and it supports the profile comparison of Figure~\ref{fig:matrix} and the fixed-profile evaluation of Section~\ref{sec:results}. A larger production set of eight sequences, $19{,}440M$ in total and 194,408 states, was produced afterwards and encoded with the selected profile, providing the production storage budget of Table~\ref{tab:long-storage} (Section~\ref{sec:results}). The two sets belong to the same GRMHD data collection.

We examine the fixed channel-aware profile at three temporal scales, all 600 non-overlapping $5M$ windows, all 60 non-overlapping $50M$ blocks, and the 18 full sequences. Each $5M$ window contains 51 states and each $50M$ block contains 501. The longer tests pool the underlying error sums before evaluating the NRMSE, rather than averaging short-window ratios, and test sustained accuracy as well as storage overhead. Windows within a trajectory are correlated and are not independent simulation realizations. 

For a matched comparison of precision choices, we also encode the first, middle, and final $5M$ window of every sequence using uniform q8, q9, q10, q12, and q16 profiles. These 54 windows are a subset of the 600-window evaluation, not additional samples. The uniform profiles are compared only on this subset. The uniform profiles apply a single bit depth and the 99.9th percentile to all primitive channels at native-MeshBlock scale. All profiles use the same input states, exact anchors, predictor, and lossless backend. The channel-aware profile follows Table~\ref{tab:profile} and the local $u$ scaling specified above, and it is held fixed across all 18 sequences and the GRRT calculations. 

All cadence comparisons follow the same densely calculated trajectory. The original $0.1M$ sequence serves as the reference, and the sparse $0.2M$, $0.5M$, and $1M$ baselines retain every second, fifth, and tenth state. For the field-level test, missing states are reconstructed by linear interpolation of primitive variables between adjacent states in physical time, as in the temporal interpolation used by \ipole.

All storage and field comparisons use the same primitive-only content, $\rho$, $u$, $\widetilde{u}^i$, and $B^i$. The original baselines retain their native gzip-compressed HDF5 representation. The DDC total includes all required containers, indices, and unchanged exact anchors.

\subsection{Field and storage criteria}

For each reconstructed quantity $f$, we use the original $0.1M$ sequence as the reference and define the normalized root-mean-square error as
\begin{equation}
\operatorname{NRMSE}(f) =
\left[
\frac{\sum_i (f_i-f_{{\rm ref},i})^2}
{\sum_i f_{{\rm ref},i}^2}
\right]^{1/2},
\label{eq:nrmse}
\end{equation}
where the sum extends over all spatial elements and all common non-anchor
evaluation times within a test interval. This is the relative $L_2$ error. Zero indicates exact agreement, whereas
unity means that the reconstruction-error norm equals the reference-field
norm.

The evaluated set, denoted by $\mathcal{F}$, includes $\rho$, $u$, all three components of $\widetilde{u}^i$ and $B^i$, and four quantities calculated from the primitive component arrays. Let
$\ell_\epsilon(x)=\ln[\max(x,\epsilon)]$, with $\epsilon=10^{-30}$. We first form the Euclidean component sums
\begin{equation}
  v_{\rm comp}^2 = \sum_{i=1}^{3}(\widetilde{u}^i)^2,
  \qquad
  B_{\rm comp}^2 = \sum_{i=1}^{3}(B^i)^2.
  \label{eq:component-sums}
\end{equation}
The velocity diagnostic is $v_{\rm comp}^2$, while $B_{\rm comp}^2$ is an auxiliary sum used to define the three logarithmic diagnostic proxies
\begin{equation}
\begin{aligned}
  m_B &= \ell_\epsilon(B_{\rm comp}^2/2), \\
  m_\sigma &= \ell_\epsilon(B_{\rm comp}^2)-\ell_\epsilon(\rho), \\
  m_\beta &= \ell_\epsilon(2u)-\ell_\epsilon(B_{\rm comp}^2).
\end{aligned}
\label{eq:derived-proxies}
\end{equation}
These component-based quantities are included as sensitive validation diagnostics. Nonlinear combinations can reveal errors that are less apparent in individual primitive components. The three $m$ quantities are logarithmic proxies, not the covariant $b^2$, magnetization, or plasma beta used by the radiative model. The floor keeps the logarithmic proxies finite. No density or funnel mask is used. Exact anchor times are excluded because their codec error is zero by construction.

We assess reconstruction accuracy relative to a practical sparse-output baseline. For every $f\in\mathcal{F}$, we calculate one NRMSE for the decoded $0.1M$ \ddc\ sequence and another for linear interpolation of the original $0.5M$ sequence. Their ratio is
\begin{equation}
  R_f =
  \frac{\operatorname{NRMSE}_{\mathrm{DDC}}(f)}
       {\operatorname{NRMSE}_{0.5M}(f)} .
  \label{eq:field-error-ratio}
\end{equation}
Because both NRMSE values use the same original $0.1M$ reference, $R_f<1$ favors \ddc, $R_f=1$ indicates equal errors, and $R_f>1$ favors sparse interpolation.

Good performance in most variables should not conceal a large error in one physically important quantity. We therefore characterize each test interval by the least favorable field-error ratio,
\begin{equation}
R_{\rm worst} = \max_{f\in\mathcal{F}} R_f .
\label{eq:worst-error-ratio}
\end{equation}
The condition $R_{\rm worst}<1$ guarantees that every evaluated quantity has a smaller NRMSE with the $0.1M$ \ddc\ sequence than with $0.5M$ interpolation. This is a relative performance measure rather than an absolute reconstruction error. For example, $R_{\rm worst}=0.8$ means that even the least favorable evaluated quantity has only 80\% of the NRMSE produced by the interpolation baseline.

Storage performance is evaluated independently using
\begin{equation}
R_{\rm storage} =
\frac{B_{\mathrm{DDC}}}{B_{0.5M}},
\label{eq:storage-ratio}
\end{equation}
where $B_{\mathrm{DDC}}$ is the total number of bytes required to reconstruct the $0.1M$ \ddc\ sequence, including the encoded intermediate states, exact anchors, and indexing metadata. $B_{0.5M}$ is the size of the matching original-data sequence saved every $0.5M$ with the same variables.
Thus, $R_{\rm storage}<1$ means that the $0.1M$ \ddc\ representation occupies less space than the sparse original-data baseline. For example, $R_{\rm storage}=0.4$ corresponds to using 40\% of the baseline storage, or a 60\% reduction.

All measured $B_{\mathrm{DDC}}$ byte counts refer to the committed $0.1M$ archive after encoding.
They exclude restart checkpoints and the transient dense non-anchor files used by the present asynchronous implementation. The denominator is likewise an archive-retention comparison and does not include restart products.

A test interval satisfies the joint storage--accuracy criterion only when
\begin{equation}
R_{\rm storage}<1
\qquad\mathrm{and}\qquad
R_{\rm worst}<1 .
\label{eq:joint-criterion}
\end{equation}
The first condition requires the $0.1M$ \ddc\ sequence to be smaller than the original $0.5M$ sequence, while the second requires it to be more accurate than $0.5M$ interpolation for every evaluated quantity.

\subsection{Polarized slow-light GRRT setup}

We use polarized GRRT to test the effects of fluid-data compression and temporal cadence under controlled conditions, and separately measure the cost of native \ddc\ access for M87*.
Slow-light GRRT has been implemented in RAPTOR, \ipole, and Blacklight \citep{bronzwaer2018raptor,moscibrodzka2021circular,white2022blacklight}. These codes solve the covariant transfer problem \citep{younsi2012grrt}. The CoportSL framework reduces the cost of slow-light calculations by applying full slow-light treatment only where the emission, absorption, and Faraday contributions require it \citep{Zhou:2026sew}. We use \kpolaris\ \citep{zhang_2026_22879728}, an open-source GPU-accelerated polarized radiative transfer code that combines parallel ray integration with snapshot streaming for efficient slow-light calculations.

The slow-light tests use the SANE and MAD sequences at $a_*=+0.5$. For each model, we calculate 16 matched $256^2$ images at observer times $26196M$ to $26203.5M$ in $0.5M$ steps, separately at 86, 230, and 345~GHz. Each calculation is repeated with original $0.1M$ input, reconstructed $0.1M$ DDC input, and original $0.2M$, $0.5M$, and $1M$ input. This gives 96 matched five-way comparisons and 480 full-Stokes images. These simulations use the same fluid trajectory, observer times, GRRT solver, and radiative settings across all five inputs.

The M87* scaling uses $6.5\times10^9\,M_\odot$, a source distance of 16.9~Mpc, inclination $17^\circ$, and a $160\,\mu{\rm as}$ square field of view. Both the SANE and MAD normalizations match a $0.5$~Jy target at 230~GHz and are held fixed across all input representations and frequencies.
We adopt the $R_{\rm high}$ electron-temperature prescription \citep{moscibrodzka2016m87}, with $R_{\rm high}=40$ for MAD and $R_{\rm high}=1$ for SANE. For an isotropic thermal Maxwell--J\"uttner electron distribution, we compute full-Stokes thermal synchrotron emission and absorption with the Dexter fitting formulae, together with the associated thermal Faraday rotation and conversion coefficients \citep{dexter2016grtrans}. Cells with $\sigma>1$ contribute no emission or transfer coefficients.

We compare all four pixel-resolved Stokes channels. For each model and frequency, image NRMSE is evaluated jointly over pixels and all 16 times. Each image is also checked separately. Integrated Stokes $I$ NRMSE uses the 16 image-integrated values as a scalar statistic. Retaining full Stokes information is useful because circular polarization and Faraday conversion probe magnetic structure absent from total intensity alone \citep{moscibrodzka2021circular}.

The performance benchmark repeats both models and all three frequencies at $512^2$, using 16 observer times near the same interval. Each group includes fast light, original $0.5M$ slow light, and two $0.1M$ DDC configurations, for another 384 images. Two additional original $0.1M$ truth images, one per model at the middle sampled time and 230~GHz, check codec fidelity at this resolution. They are not a full 16-time reference series. The required fluid-time span is about $264M$, and the 16-image batch selects 2742 dense states, including time-window guards, from each DDC sequence.

The reported timings use one NVIDIA RTX 5090. Each slow-light job batches 16 observer times, and the fast-light baseline uses 16 individual snapshot jobs. Bulk original $0.5M$ inputs, exact anchors, and the disposable DDC spatial cache are on local NVMe storage. Both slow-light paths use a configurable fluid-state cache. Reported times include process startup and image output but exclude the one-time staging and working-cache preparation. These are workflow timings, not an isolated comparison of compression kernels or identically optimized file readers.

\section{Results}
\label{sec:results}

\begin{figure*}[tbp!]
\centering
\includegraphics[width=\textwidth]{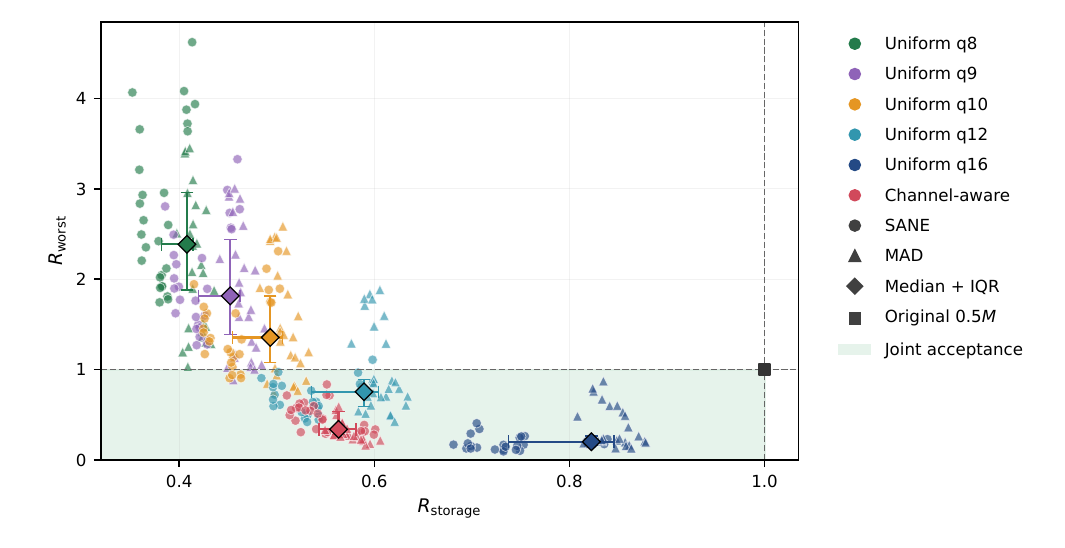}
\caption{Rate--distortion for 54 matched $5M$ windows. Colors denote precision profiles. Circles and triangles denote SANE and MAD. Diamonds and bars show medians and interquartile ranges. The shaded region satisfies both criteria.}
\label{fig:matrix}
\end{figure*}

\begin{table*}[tbp!]
\centering
\caption{Storage of the eight production sequences.}
\label{tab:long-storage}
\small\setlength{\tabcolsep}{4pt}
\begin{tabular}{lrrrrrrr}
\toprule
Case & Duration & States & \multicolumn{3}{c}{Storage (TiB)} & Reduction & Compression \\
\cmidrule(lr){4-6}
 & ($M$) & & Original $0.1M$ & Original $0.5M$ & DDC 0.1M & vs. $0.5M$ & vs. $0.1M$ \\
\midrule
SANE, $a_*=-0.94$ & 1,450 & 14,501 & 1.900 & 0.380 & 0.214 & 43.8\% & 8.90$\times$ \\
SANE, $a_*=0$ & 1,450 & 14,501 & 1.896 & 0.379 & 0.181 & 52.3\% & 10.47$\times$ \\
SANE, $a_*=+0.5$ & 5,350 & 53,501 & 6.998 & 1.400 & 0.622 & 55.6\% & 11.25$\times$ \\
SANE, $a_*=+0.94$ & 1,470 & 14,701 & 1.925 & 0.385 & 0.192 & 50.2\% & 10.04$\times$ \\
MAD, $a_*=-0.94$ & 1,450 & 14,501 & 1.895 & 0.379 & 0.197 & 47.9\% & 9.60$\times$ \\
MAD, $a_*=0$ & 1,450 & 14,501 & 1.887 & 0.378 & 0.175 & 53.7\% & 10.79$\times$ \\
MAD, $a_*=+0.5$ & 5,350 & 53,501 & 6.965 & 1.393 & 0.769 & 44.8\% & 9.06$\times$ \\
MAD, $a_*=+0.94$ & 1,470 & 14,701 & 1.916 & 0.383 & 0.205 & 46.6\% & 9.36$\times$ \\
\midrule
All sequences & 19,440 & 194,408 & 25.383 & 5.077 & 2.554 & 49.7\% & 9.94$\times$ \\
\bottomrule
\end{tabular}
\end{table*}

This section reports the results of the evaluation plan set out in Section~\ref{sec:evaluation}. We first select the quantization profile, validate it across all GRMHD configurations, and report the storage achieved on the production set. We then test the fidelity of polarized slow-light images, and finally measure the cost of native \ddc\ access.

\subsection{Storage--accuracy across GRMHD configurations}

We assess each profile using $R_{\rm storage}$ and $R_{\rm worst}$ from Section~\ref{sec:evaluation}, both of which must be below unity. Figure~\ref{fig:matrix} shows the $R_{\rm storage}$--$R_{\rm worst}$ plane for the 54 matched windows. All five uniform profiles satisfy the storage condition in every window, so acceptance depends on accuracy alone. q8--q10 fail most windows, q12 fails ten windows (all limited by the logarithmic magnetic-pressure proxy), and only q16 passes all 54 windows, with a largest $R_{\rm worst}$ of 0.873.

The uniform profiles, however, apply the same bit depth to every channel. The channel-aware profile instead assigns bit depths per channel (Table~\ref{tab:profile}) and scales $u$ locally. It passes all 54 windows with a median $R_{\rm storage}$ of 0.563. At the median of the 54 paired comparisons it uses 29.3\% fewer stored bytes than uniform q16. Uniform q16 is more accurate on average (median $R_{\rm worst}$ 0.201 against 0.339), but neither profile is more accurate in every window, and the channel-aware profile meets the same acceptance criterion at lower storage cost.

Channel-aware encoding is not simply a matter of bit allocation. With the same channel bit depths but a single $u$ scale per native MeshBlock, only 48 of the 54 windows pass. The finer tiles keep small thermal residuals from inheriting the scale of more strongly varying regions in the same MeshBlock. All subsequent results use the fixed channel-aware profile.

Over the entire test set, the fixed profile passes 600/600 $5M$ windows, 60/60 $50M$ blocks, and all 18 full sequences. Over the 600 windows the median $R_{\rm storage}$ is 0.573 and the median $R_{\rm worst}$ is 0.328. The largest $R_{\rm worst}$ is 0.859, set by $\rho$ in the SANE $a_*=+0.5$ window near $27120M$, where the DDC and original-$0.5M$ NRMSE values are $6.02\times10^{-4}$ and $7.01\times10^{-4}$.

For the $50M$ blocks the median and maximum $R_{\rm worst}$ are 0.358 and 0.717, and for the full sequences they are 0.355 and 0.610. Longer pooling does not conceal a failing short window, since every constituent $5M$ window also passes. All 1200 completed GOPs of the test set pass the recorded integrity checks. These measurements test transfer across epochs and magnetic states, but windows from the same trajectory are not independent statistical samples.

Table~\ref{tab:long-storage} summarizes the storage of the eight production sequences, which were encoded with the selected profile.
The committed DDC archives total 2.554~TiB, compared with 25.383~TiB for the dense original inventory and 5.077~TiB at $0.5M$. This corresponds to a 9.94-fold dense compression and a 49.7\% reduction relative to sparse storage. Unchanged exact anchors remain available for conventional analysis.

\subsection{Slow-light fidelity}

For each model, frequency, and Stokes channel, we evaluate one image NRMSE jointly over all pixels in all 16 matched images. Table~\ref{tab:slowlight-fidelity} divides this aggregate DDC NRMSE by the aggregate original-data NRMSE at the cadence specified in each row, and it does not average 16 individual ratios. All calculations use original $0.1M$ slow light as the reference.

\begin{table}[tbp!]
\centering
\caption{NRMSE ratios of $0.1M$ DDC to original input at cadence $\Delta t_{\rm orig}$. Image columns pool 16 times. $I_{\rm int}$ uses their integrated intensities. Max. is the largest individual Stokes-image ratio.}
\label{tab:slowlight-fidelity}
\fontsize{8}{9.5}\selectfont\setlength{\tabcolsep}{1.8pt}
\begin{tabular}{llrrrrrrr}
\toprule
Model & GHz & $\Delta t_{\rm orig}/M$ & $I$ & $Q$ & $U$ & $V$ & $I_{\rm int}$ & Max. \\
\midrule
MAD & 86 & 0.2 & 1.411 & 2.454 & 2.673 & 2.123 & 1.968 & 4.479 \\
 &  & 0.5 & 0.430 & 0.394 & 0.406 & 0.353 & 0.278 & 0.588 \\
 &  & 1 & 0.164 & 0.107 & 0.109 & 0.096 & 0.071 & 0.198 \\
\addlinespace[2pt]
MAD & 230 & 0.2 & 1.086 & 1.149 & 1.128 & 1.490 & 1.118 & 2.658 \\
 &  & 0.5 & 0.360 & 0.308 & 0.292 & 0.355 & 0.172 & 0.536 \\
 &  & 1 & 0.159 & 0.098 & 0.095 & 0.109 & 0.046 & 0.204 \\
\addlinespace[2pt]
MAD & 345 & 0.2 & 1.005 & 1.059 & 1.027 & 1.252 & 1.169 & 2.763 \\
 &  & 0.5 & 0.340 & 0.334 & 0.328 & 0.463 & 0.261 & 0.961 \\
 &  & 1 & 0.158 & 0.135 & 0.129 & 0.169 & 0.080 & 0.375 \\
\addlinespace[2pt]
SANE & 86 & 0.2 & 3.097 & 3.122 & 3.147 & 3.541 & 0.192 & 3.812 \\
 &  & 0.5 & 0.391 & 0.396 & 0.396 & 0.453 & 0.025 & 0.496 \\
 &  & 1 & 0.097 & 0.098 & 0.098 & 0.112 & 0.006 & 0.121 \\
\addlinespace[2pt]
SANE & 230 & 0.2 & 2.093 & 2.106 & 2.267 & 2.127 & 0.289 & 2.645 \\
 &  & 0.5 & 0.266 & 0.267 & 0.287 & 0.268 & 0.037 & 0.339 \\
 &  & 1 & 0.067 & 0.067 & 0.073 & 0.068 & 0.010 & 0.082 \\
\addlinespace[2pt]
SANE & 345 & 0.2 & 1.771 & 1.725 & 1.917 & 1.693 & 0.274 & 2.212 \\
 &  & 0.5 & 0.224 & 0.218 & 0.242 & 0.214 & 0.035 & 0.283 \\
 &  & 1 & 0.057 & 0.055 & 0.061 & 0.054 & 0.009 & 0.071 \\
\bottomrule
\end{tabular}
\end{table}

Against the original $0.5M$ baseline, the ratios across the six model--frequency groups and four Stokes channels span $0.214--0.463$, corresponding to error reductions of $53.7--78.6\%$. Every one of the $384$ individual Stokes-image comparisons improves over the $0.5M$ baseline, including $86$~GHz, and the largest individual ratio is $0.961$.

\begin{figure*}[tbp!]
\centering
\includegraphics[width=\textwidth]{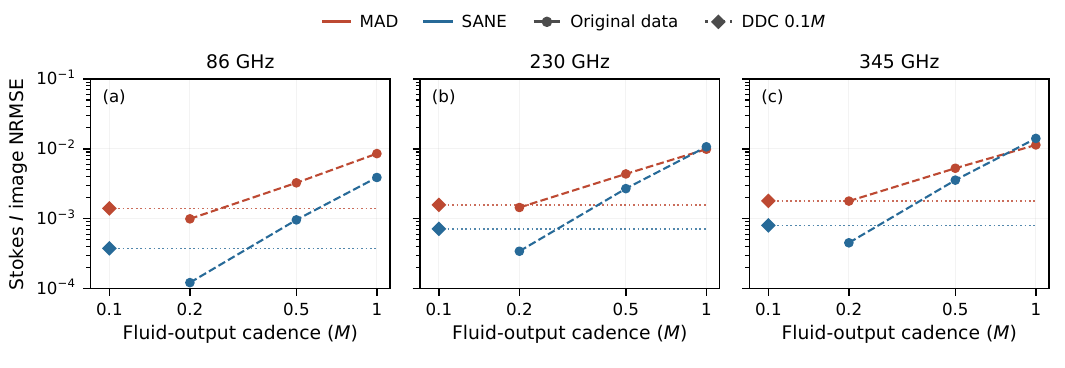}
\caption{Stokes $I$ image NRMSE versus fluid-output cadence for MAD ($R_{\rm high}=40$) and SANE ($R_{\rm high}=1$) at three frequencies. Circles denote original data. Diamonds and dotted lines mark the fixed $0.1M$ DDC result.}
\label{fig:cadence}
\end{figure*}

The denser $0.2M$ and sparser $1M$ baselines bracket this result. The $0.2M$ and $1M$ comparisons give aggregate image-NRMSE ratios of 1.005--3.541 and 0.054--0.169, respectively. All 384 individual Stokes-image comparisons improve against original $1M$. The largest individual ratios are 4.479 against $0.2M$ and 0.375 against $1M$. Thus, the image-level accuracy of the $0.1M$ \ddc\ input exceeds the sparser $0.5M$ and $1M$ baselines in every comparison, while the original $0.2M$ input is more accurate than the \ddc\ reconstruction. For MAD, the \ddc\ reconstruction reaches comparable accuracy while using about 80\% less storage than the original $0.2M$ output. For SANE, it does not reach the original $0.2M$ level, a gap that may reflect the turbulent structure and variability of the SANE flow.

Integrated Stokes $I$ ratios improve uniformly against the original $0.5M$ baseline, spanning 0.025--0.278 across the six groups. Against the denser $0.2M$ input the ratios are mixed, with the three MAD groups above unity (1.118--1.968) and the three SANE groups below it (0.192--0.289). All six groups improve against $1M$. Smaller pixel-wise errors need not produce a smaller integrated error, because signed residuals can cancel when summed over an image.

\begin{figure*}[tbp!]
\centering
\includegraphics[width=\textwidth]{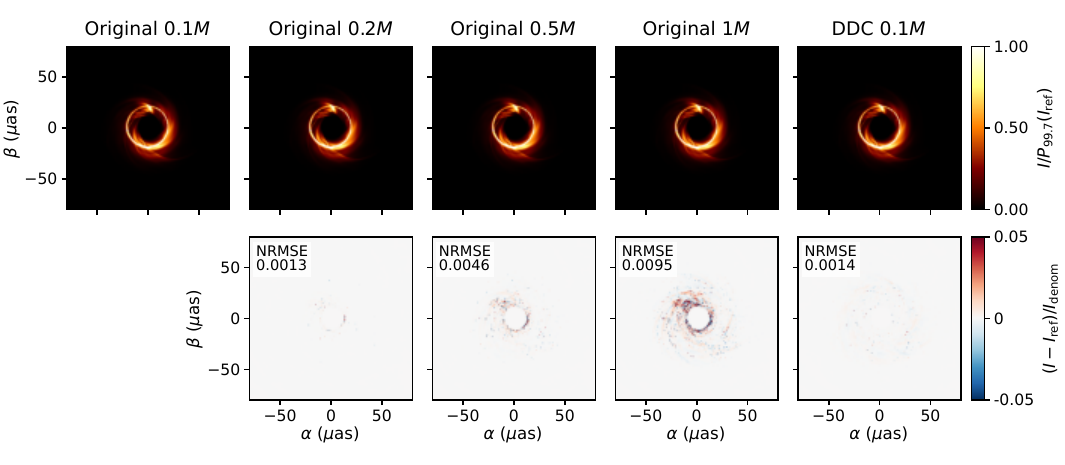}
\caption{Matched $256^2$ Stokes $I$ images at 230~GHz for MAD $a_*=+0.5$. Columns show original $0.1M$, $0.2M$, $0.5M$, $1M$, and DDC $0.1M$ input. The upper panels show $I/P_{99.7}(I_{\rm ref})$. The lower panels show $(I-I_{\rm ref})/I_{\rm denom}$ with $I_{\rm denom}=\max(|I_{\rm ref}|,0.01I_{\rm ref,max})$ on a common range of $[-0.05,0.05]$. Each lower panel is labeled with its NRMSE relative to the original $0.1M$ reference, as a fraction. Values outside either range are clipped.}
\label{fig:representative}
\end{figure*}

Figure~\ref{fig:cadence} shows the Stokes $I$ image NRMSE versus fluid-output cadence at all three frequencies. The original $0.1M$ reference has zero error against itself and is omitted from the logarithmic error axes. Dotted lines mark the fixed $0.1M$ \ddc\ error, not a test at a different cadence. Figure~\ref{fig:representative} shows the MAD $a_*=+0.5$ images at 230~GHz from the middle observer time, $t_{\rm obs}=26200M$. Each lower panel is labeled with its NRMSE relative to the original $0.1M$ reference. The labeled differences decrease monotonically as the input cadence becomes denser, and the \ddc\ reconstruction gives $0.0014$, between the $0.0013$ of the $0.2M$ input and the $0.0046$--$0.0095$ of the sparser inputs.

\subsection[Native-reader performance]{Native-reader performance}

\begin{table*}[tbp!]
\centering
\caption{Workflow timings at $512^2$ on an Nvidia RTX 5090. Ranges cover six model--frequency trials, each producing 16 images. Times include process startup but exclude cache preparation.}
\label{tab:grrt-performance}
\small\setlength{\tabcolsep}{5pt}
\begin{tabular}{lrrr}
\toprule
Workflow & Selected fluid states & Seconds per image & Peak GPU memory (GiB) \\
\midrule
Fast light (individual) & 16 & 16.9--17.4 & 1.28 \\
Original $0.5M$ slow light & 550 & 14.3--14.5 & 26.40 \\
DDC $0.1M$ slow light & 2742 & 12.3--13.4 & 27.03 \\
DDC $0.1M$ slow light (low memory) & 2742 & 16.5--17.0 & 7.28 \\
\bottomrule
\end{tabular}
\end{table*}

Table~\ref{tab:grrt-performance} summarizes the computational costs of the six $512^2$ model--frequency trials. The \ddc\ batches select 2742 dense states, and the \kpolaris\ slow-light workflow streams them through a bounded resident window rather than loading the full sequence. The retarded coordinate time along each ray selects the neighboring states. 
In the optimized configuration, the workflow takes 12.3--13.4~s per image, versus 16.9--17.4~s for fast light and 14.3--14.5~s for original $0.5M$ slow light, about 1.3 and 1.1 times faster, respectively, while accessing five times more densely sampled evolution. Part of this difference comes from batching. The fast-light baseline runs 16 independent snapshot jobs, while the slow-light paths process all 16 observer times in a single job, reducing the per-image overhead. Peak GPU memory is about 27.03~GiB in the optimized configuration, compared with 26.40~GiB for original $0.5M$ slow light and 1.28~GiB for fast light. The memory footprint is set by the number of cached files rather than by the length of the sequence, so the reported values do not represent a minimum. A low-memory configuration runs the same \ddc\ workflow in 7.28~GiB of GPU memory at 16.5--17.0~s per image, bringing dense slow-light calculations within reach of entry-level GPUs at the cost of a longer per-image time.

\section{Summary}
\label{sec:summary}

We have presented the Dense Dump Codec (\ddc), an error-controlled compression method for dense GRMHD time series. DDC predicts intermediate simulation states from two exact anchor states using their physical times, then stores the prediction residuals with channel-dependent quantization. The anchors remain unchanged, while the residuals preserve the intervening evolution calculated by the simulation. Indexed chunks provide access to selected times and variables without decoding the full sequence. This design combines temporal prediction, controlled representation errors, and selective decoding in a format suited to slow-light radiative transfer.

Validation on the test set shows that the selected channel-aware profile satisfies the joint storage--accuracy criterion in all 600 short windows, 60 $50M$ blocks, and 18 full sequences. On the eight production sequences, $19{,}440M$ in total and 194,408 states, the committed archives are 9.94 times smaller than the original dense output in aggregate and use 49.7\% less storage than the original $0.5M$ baseline, while preserving exact analysis anchors. In the polarized slow-light tests, the aggregate DDC-to-$0.5M$ image-NRMSE ratios span 0.214--0.463 across the three frequencies, with improvement in all 384 individual Stokes-image comparisons. All image-level comparisons also improve against the original $1M$ input, while the original $0.2M$ input remains comparable or more accurate. Integrated-intensity errors are smaller than at $0.5M$ in every group, but only the SANE groups improve on the $0.2M$ input. At a fixed storage budget, keeping more calculated states with controlled representation error can therefore be preferable to discarding states and later interpolating the turbulent evolution.

In GPU-accelerated \kpolaris code tests, a native 16-image batch at $512^2$ takes about 12.3--13.4~s per image on an RTX 5090, about 1.3 times faster than the measured fast-light cost, and a low-memory configuration runs the same workflow in 7.28~GiB of GPU memory, within reach of entry-level GPUs. Together, the storage, accuracy, and timing results show that slow-light radiative transfer at $0.1M$ cadence is practical on a single GPU. Since storing the dense evolution is no longer the limiting cost, the output cadence of a simulation can be chosen from the variability that an analysis must resolve rather than from the storage budget. The same container and reader can be carried over when the method is applied to other simulation codes.

More broadly, \ddc\ provides a compact, solver-independent representation of densely sampled GRMHD evolution for slow-light radiative transfer and other time-dependent analyses. This allows the temporal resolution of the archived data to be retained without tying the format to a particular GRRT code. However, the compression profile validated here should be regarded as a tested operating point rather than a universal precision prescription. For different simulations or scientific objectives, the effects of compression should therefore be reassessed at both the level of the reconstructed fields and that of the target observables. In future work, we will extend this validation to a broader range of simulations and incorporate residual encoding directly into GRMHD output.

\section*{Data and Software}
The Dense Dump Codec is publicly available under the BSD-3-Clause
license at \url{https://github.com/zelinzh/dense-dump-codec}.
The repository includes documentation and examples for encoding,
decoding, and validating simulation sequences. An archived software
release is available through Zenodo \citep{zhang_2026_22845273}.

\begin{acknowledgments}
The work is partly supported by NSFC Grant No. 12275004, 12547132, 12547127, and No. 12588101. Zhenyu Zhang is also supported by the Postdoctoral Science Preferential Funding of Zhejiang Province under Grant No. ZJ2026086.
\end{acknowledgments}

\software{Dense Dump Codec \citep{zhang_2026_22845273}, KHARMA \citep{prather2024kharma},
KPolaris \citep{zhang_2026_22879728}}

\clearpage
\bibliographystyle{aasjournal}
\bibliography{references}

\end{document}